\documentstyle[12pt]{article}
\begin{document}
\baselineskip=15pt
%
\newcommand{\bc}{\begin{center}}
\newcommand{\ec}{\end{center}}
\newcommand{\be}{\begin{equation}}
\newcommand{\ee}{\end{equation}}
\newcommand{\PP}{\mbox{$\Psi$}}
\newcommand{\wdg}{\mbox{$\wedge$}}
\newcommand{\bq}{\begin{eqnarray}}
\newcommand{\eq}{\end{eqnarray}}
\newcommand{\del}{\mbox{$\delta$}}
\newcommand{\AC}{\mbox{${\bf {\cal A}}$}}
\newcommand{\BA}{\mbox{${\bf A}$}}
\newcommand{\BB}{\mbox{${\bf B}$}}
\newcommand{\DC}{\mbox{${\bf {\cal D}}$}}
\newcommand{\xx}{\mbox{${\bf x}$}}
\newcommand{\yy}{\mbox{${\bf y}$}}
\newcommand{\Lam}{\mbox{${\Lambda}$}}
\newcommand{\lam}{\mbox{${\lambda}$}}
\newcommand{\eps}{\mbox{$\epsilon$}}
\newcommand{\vol}{\mbox{$d^{3}{\bf x}$}}
\newcommand{\Sc}{Schr\"odinger\ }
\newcommand{\p}{\mbox{$\varphi$}}
\newcommand{\der}{\partial}
\begin{titlepage}
\bigskip
\bc
{\Large \bf Constructing the Leading Order Terms of the Yang-Mills \Sc 
Functional} 
\ec
\vskip1in
\bc
{\large
Paul Mansfield and Marcos Sampaio\footnote{Supported by CNPq - Conselho 
Nacional de Desenvolvimento 
Cient\'{\i}fico e Tecnol\'{o}gico - Brasil }
\vskip2.0cm
Department of Mathematical Sciences

University of Durham

Durham, DH1 3LE, England}

{\it P.R.W.Mansfield@durham.ac.uk}
{\it M.D.R.Sampaio@durham.ac.uk}

\ec
\vskip1cm
\begin{abstract}
\noindent
The leading short-time behaviour of the Yang Mills \Sc functional is obtained 
within a local expansion in the fields.

\end{abstract}

\end{titlepage}

\section{\bf Introduction}

In the Schr\"odinger representation the vacuum of a quantum field theory is a 
functional of the eigenvalue, $\varphi(x)$, of the field on a fixed,
spacelike, quantisation surface with co-ordinate $x$. In perturbation theory
its logarithm, $W[\varphi]$, is a sum of connected Feynman diagrams, and is, 
in general, a non-local functional. If $\varphi$ varies very slowly on a
length scale of the inverse of the lightest mass in the theory, then  
$W[\varphi]$ reduces to a sum of local functionals, that is to say, 
integrals of products of $\varphi$ and a finite number of its derivatives
at the same spatial point. As it stands this observation
appears to have little physical interest since we need more rapidly
varying  fields to probe the internal structure of particles which occurs  
on shorter scales. However, when $W$ is evaluated for a field that is
obtained from $\varphi$ by a scale transformation in which $x\rightarrow 
\lambda x$ then it is analytic in $\lambda$. This may be used to
reconstruct $W[\varphi]$ for arbitrary $\varphi$
from the local expansion \cite{Paul2}. Furthermore, the coefficents of the 
expansion are determined by the Schr\"odinger equation which may be
contructed  so as to act directly on this expansion, again by exploiting 
analyticity \cite{Paul}. For theories that are classically massive this local 
expansion appears already within the framework of standard semi-classical 
perturbation theory, but for Yang-Mills theory, which is classically massless, 
 the leading order contribution to $W$ does not reduce to a local
expansion for slowly varying fields. Nonetheless quantum effects generate 
a non-zero mass-gap, and so the full expression for $W$ does have such an 
expansion, as has been seen in Monte-Carlo simulations of lattice gauge
theory \cite{Jef2,Jef3,ari}. In principle this will be determined by
solving the Schr\"odinger equation, but in practice the construction of this 
equation to a sufficent order to generate reliable results is some way
off. It  would have been useful to study this local expansion using
standard semi-classical techniques. As we cannot we will study instead the
so-called Schr\"odinger functional, which is just the matrix element of the
Euclidean time-evolution operator for time $\tau$, between eigenstates of  
the gauge-field $\bf A$.

\be \Phi_{\tau}[{\bf A}, {\bf A}']\ =
\langle {\bf A }|  e^{-{\em H}  \tau /\hbar} | {\bf A}' \rangle .
\label{ME}
\ee
(Here $\em H$ is the Hamiltonian ${\em H}[{\bf A},{\bf E}]= -\frac{1}{2g^{2}} 
\int d^{3} {\bf x}\ {\it tr} ({\bf B}^2 + {\bf E}^2)$, where ${\bf E}=-\dot{{\b
f A}}$, ${\bf B}= \nabla \wdg{\bf A}+{\bf
A}\wdg {\bf A}$, ${\bf A}={\bf A}^{A} T^{A}$,  ${\it tr}(T^A T^B)= - \del^{AB}$
, $[T^{A},T^{B}]= f^{ABC} T^{C}$ and we work in the Weyl gauge, $A_0 = 0$.)

Again the logarithm of this functional,${W_{\tau}[\bf A, \bf {A}']}$ is a 
sum of connected diagrams, and is non-local, but having introduced the
length-scale $\tau$ into the problem results in a local expansion 
for fields that vary slowly, even within the semi-classical expansion. In 
other words we can solve for the short-time behaviour using both a local
expansion, and semi-classical perturbation theory. This enables us to
compare the efficacy of solving the problem in two different ways, i.e. by
solving the Hamilton-Jacobi equation for the derivative 
expansion and by evaluating the leading order contribution to the
functional  integral. Furthermore the result of the semi-classical calculation 
is useful as it will still be a good leading order approximation to the full 
expression for times that are small in comparison to the inverse of the
lightest glueball mass. The \Sc functional has been studied in lattice QCD
\cite{Luscher} as a functional integral over a space-time with boundaries 
given by the quantisation surfaces at times $0$ and $\tau$. The
divergences have been studied at one-loop, where it was found that they
could be cancelled by quark counter-terms on the space-time boundary
\cite{Sint,Sint2}.

\bigskip
     We begin by considering the differential equation approach to 
${W_{\tau}[\bf A, \bf {A}']}$. Now $\Phi_\tau$ satisfies the \Sc equation
with initial condition
\be
-\hbar{\partial\over\partial \tau}\Phi_{\tau}[{\bf A}, {\bf A}']=H\, 
\Phi_{\tau}[\bf A, \bf {A}'],
\quad\lim_{{\bf \tau} \rightarrow 0}\ \Phi_{\bf \tau} [{\bf A }, {\bf A}'] = 
\del [\bf A - {\bf A}']
\ee
In the Schr\"{o}dinger representation the Yang-Mills electric field is  
represented by ${\hat {E}^{\mu}_{C}} ({\bf x}) = {\it i}\hbar g^2\
{{\del}\over {{\del}  A^{C}_{\mu}({\bf x})}}$, so that 

\be
{\em H} = \left ( - \frac{1}{2} \hbar g^2\Delta + g^{-2}\cal B \right ) ,\ 
\Delta \ {\equiv} \int d^{3}{\bf x}\ \frac{\del}{\del
{\bf {A}^{C}}({\bf x})} \cdot \frac {\del}{\del {\bf {A}^{C}}(\bf x)}\ $$and$$ 
\ {\cal B} = - {\frac {1}{2}} \int d^{3}{\bf x}\
{\it tr}{\bf {B}}^{2} \ .
\label{LAPLA}
\ee
The \Sc equation (\ref{LAPLA})  must be regularized as the kinetic term  
$\Delta$ contains two functional derivatives acting at the same point of space,
 however this will not affect the leading order calculation, since if we  
set ${W_{\tau}[\bf A, \bf {A}']}={w[\bf A, \bf {A}']}/(\hbar g^2)$ 
the \Sc equation reads: 
\be
\frac{1}{2}\ \hbar g^2\ \Delta w \ - \frac{1}{2} \left( \int d^{3}{\bf x}\ 
\frac{\del w}{\del {\bf A}^{C}} \cdot \frac{\del w}{\del {\bf A}^{C}} \right) +
 {\cal {B}} + \frac{\der w}{\der t} = 0 \ .
\label{SEW}
\ee
Neglecting  the $O(\hbar g^2)$ term leads to the Hamilton-Jacobi equation
which we now solve in a derivative expansion subject to the conditions
that $w[{\bf A},{\bf A}']$ be real and invariant under simultaneous 
time-independent gauge transformations of ${\bf A}$ and ${\bf A}'$ and that 
$w[{\bf A}, {\bf A}'] = w[{\bf A}' , \bf A ]$. If we order the local expansion
of $w$ according to the mass dimension of local functionals, then the 
first gauge invariant term is $\int d^{3}{\bf x}\ {\it tr}{\AC}^{2}$, 
where ${\AC} \equiv  {\bf A} - {\bf A}'$. since the only length-scale in this
classical problem is $\tau$ this enters $w$ multiplied by $1/\tau$. This term 
will enable us to fit the initial condition since as $\tau$ becomes small 
$\exp \tau^{-1}\int d^{3}{\bf x}\ {\it tr}{\AC}^{2}\sim\del [\bf A - {\bf
A}']$.
Next we must include dimension four fields, for example 
$\int d^{3}{\bf x}\ {\it tr} {\bf B}^{2}$ which is needed to cancel a 
similar term in the Hamiltonian.
The Hamilton-Jacobi equation generates cross-terms from these two functionals
of the form

\be
\left(\frac{\del}{\del A^{R}_{\mu}({\bf x})} \int d^{3}{\bf x}'\ 
\AC^{A}_{\nu}({\bf x}')\AC^{A}_{\nu}({\bf x}')\right)
\left(\frac{\del}{\del A^{R}_{\mu}({\bf x})}\ \frac{1}{2}\ \int d^{3}{\bf x}'\ 
B^{A}_{\rho}({\bf x}')B^{A}_{\rho}({\bf x}')\right)\nonumber
\ee

\be
=
\left(2\  \AC^{R}_{\mu}|_{x}\right)
\left({\cal
D}^{RA}_{\mu \rho} B^{A}_{\rho}|_{x}\right)
\ee
where
\be  
{\cal D}^{RA}_{\mu \rho} \equiv  \eps_{\mu \alpha \rho} {\em D}^{RA}_{\alpha}\ 
\ ,\  {\em D}^{RA}_{\alpha} = (\partial_{\alpha} \del^{AR}
+ f^{RSA} A^{S}_{\alpha})\ .
\ee
This is a gauge-invariant dimension four field that should in turn be 
included in the expansion of $w$, as too should all further terms generated by
repeated applications of $\int \vol \ {\cal A} \cdot \delta/{\delta {\bf A}}$.
Thus we arrive at an ansatz for the the local expansion including terms up to 
dimension four
\bq
W[\BA,\BA'] &=& \int \vol\ {\it tr}\ \{\AC^{2}|_{x} a/\tau +\ {\bf B}^{2}|_{x} 
b\tau +\ \AC \cdot \DC B|_{x} c\tau\nonumber \\ &+& \int d^{3}\yy\ \AC(\xx) 
\cdot \Lam(\xx,\yy) \cdot \AC(\yy)\ d\tau +\ {\em D}\AC
\cdot [\AC,\AC]|_{x}\ e\tau\nonumber \\ &+& [\AC,\AC]\cdot [\AC,\AC]|_{x}\ f 
\tau
\label{ANS}
\eq 
where
\bq
\Lambda^{AB}_{\mu \nu}({\bf x},{\bf y}) &=& \frac{\del}{\del A^{A}_{\mu}
({\bf x})} \frac{\del}{\del A^{B}_{\nu}({\bf y})}\  
\frac{1}{2}\ \int \vol' B^{R}_{\rho}B^{R}_{\rho}|_{x'}  \nonumber \\&=&  
({\cal D}^{AR}_{\mu \rho}{\cal D}^{RB}_{\rho \nu}| +
f^{ABR}\epsilon_{\mu \nu \rho} B^{R}_{\rho})|_{x}\  \del^{3}({\bf x} - 
{\bf y}) \nonumber \\ &=& (\DC^{AR}_{\mu \rho}\DC^{RB}_{\rho \nu} + F^{AB}_{\mu 
\nu})|_{x}\  \del^{3}(\xx - \yy)\ ;
\eq
The constants $b,c,d,e$ are related by imposing 
$w[{\bf A}, {\bf A}'] = w[{\bf A}' , \bf A ]$, which, after some algebra 
implies that
$c = -\ b$ and $e = \frac{1}{2}\ b - 3\ d$.
Finally, if we substitute (\ref{ANS}) into the Hamilton-Jacobi equation the 
remaining coefficents are determined leading to our final result for $W$:
\bq
W_{\tau}[\BA,\BA'] &=& \ {1\over 2 \hbar g^2}\int \vol\ {\it tr} \{ - 
\frac{\AC^{2}}{\tau}|_{x}\ -\ \BB^{2}|_{x}\ \tau\ +\ \AC \cdot \DC B|_{x}\ 
\tau \nonumber \\
                &-&\ \int d^{3}\yy \AC(\xx)\cdot\Lam(\xx,\yy)\cdot\AC(\yy)\ 
\frac{\tau}{3}\ +\ {\em D}\AC\cdot[\AC,\AC]|_{x}\ 
                    \frac{\tau}{2}\nonumber \\ 
                &-&\ [\AC,\AC]\cdot[\AC,\AC]|_{x}\ \frac{\tau}{10}\}\ . 
\eq

\bigskip

  We will now compute this expression using a different method, so as to 
gauge the efficiency of the above approach. The functional integral 
representation of $W_\tau[\BA_{out},\BA_{in}]$ leads to a saddle-point 
approximation in which this is given to leading order by minus the
Euclidean action 

\be
\frac{1}{2\hbar g^2} \int \vol\ dt \  ({\bf E}^{A}\cdot{\bf E}^{A} + 
{\bf B}^{A}\cdot{\bf B}^{A})|_{x}
\label{EA} 
\ee
where, in the temporal gauge, 
\be
{\bf E} = - \dot{\BA} \;\;\; and \;\;\; {\bf B} = \nabla\ \wdg\ {\BA} + 
{\BA}\ \wdg\ {\BA}\ .
\label{EAC} 
\ee
The fields ${\BA}$ are required to be on-shell, i.e. they
satisfy the Euler Lagrange equations
\be
\ddot{\BA} = -\ D\ \wdg\ {\bf B}\;\;\ , \;\;\ D\ \wdg\ {\bf B} = \nabla\ \wdg\ 
{\bf B} + \BA\ \wdg\ {\bf B} + {\bf B}\ \wdg\  
\BA
\ee
subject to the  
boundary conditions that at 
$\BA(\xx,0) = \BA_{in}$ and  
$\BA(\xx,\tau) = \BA_{out}$. If we expand the
field as a power series in $\tau$:
$\BA(\xx,t) = \sum_{n = 0}^{\infty}\ \BA_{n}(\xx)\ \tau^{n}$
and substitute into the equations of motion and boundary conditions
we eventually obtain

\bq
\BA_{0}  &=& \BA_{in}\nonumber\\
\BA_{1} &=& \frac{\AC}{\tau}\ -\ \BA_{2}\ \tau\ -\ \BA_{3}\ \tau^{2}\ -\ 
\BA_{4}\ \tau^{3}\ -\ \BA_{5}\ \tau^{4}\ -\ O(\tau^{2})\ ,
\nonumber \\
\BA_{2} &=& -\ \frac{1}{2}\ D_{0}\ \wdg\ \BB_{0}\ , \nonumber \\
\BA_{3} &=& -\ \frac{1}{6\tau}\ \{\ D_{0}\ \wdg\ (D_{0}\ \wdg\ \AC)\ +\ \AC\ 
\wdg\ \BB_{0}\ +\ \BB_{0}\ \wdg\ \AC\ \}\ 
,\nonumber \\
\BA_{4} &=& -\ \frac{1}{12\tau^{2}}\ \{\ D_{0}\ \wdg\ (\AC\ \wdg\ \AC)\ +\ 
\AC\ \wdg\ (D_{0}\ \wdg\ \AC)\ +\ (
D_{0}\ \wdg\ \AC)\ \wdg\ \AC\ \}\ ,\nonumber \\
\BA_{5} &=& -\ \frac{1}{20\tau^{3}}\ \{\ \AC\ \wdg\ (\AC\ \wdg\ \AC)\ +\ 
(\AC\ \wdg\ \AC)\ \wdg\ \AC\ \}\ .
\eq
where we neglected terms of $O(L^{-n})\ n > 4$ \ and \  $D_0\ \wdg \;\;\;\ = 
\nabla\ \wdg \;\;\;\ + \BA_0\ \wdg \;\;\;\
+ \;\;\;\ \wdg\ \BA_0$. The subscript $0$ denotes that the gauge potential
has been taken as $\BA_0$, and ${\cal A}=\BA_{out}-\BA_{in}$.
Substituting this into the Euclidean action and some considerable algebra
eventually leads to the same expression as previously for $W_{\tau}[\BA,\BA']$. 

\bigskip
To conclude, we have computed the local expansion of the leading order 
term in the semi-classical expansion of the \Sc  functional for Yang-Mills 
theory using two different approaches, the first by substituting our local
expansion directly into the Hamilton-Jacobi equation, the second by
computing the on-shell Euclidean action. It turned out that the former
approach was far more efficent.
The result describes the leading short-time behaviour. As a final check on 
our results note that when the initial and final gauge-fields are 
identified ${\cal A}=0$ and our expression reduces, as it should, 
to $-\tau\int d^3{\bf x}{\bf B}^2/(2g^2)$, i.e. minus the Euclidean action
evaluated for a time-independent potential.
\vskip0.5cm
M.S. acknowledges a grant from CNPq/Brasil. 
\end{document}